# Electrical Properties of Atomic Layer Deposited Aluminum Oxide on Gallium Nitride

Michele Esposto[a)], Sriram Krishnamoorthy, Digbijoy N. Nath, Sanyam Bajaj, Ting-Hsiang Hung, and Siddharth Rajan

ECE Department, The Ohio State University, Columbus, Ohio 43210, USA

**ABSTRACT**

We report on our investigation of the electrical properties of metal/$Al_2O_3$/GaN metal-insulator-semiconductor (MIS) capacitors. We determined the conduction band offset and interface charge density of the alumina/GaN interface by analyzing capacitance-voltage characteristics of atomic layer deposited $Al_2O_3$ films on GaN substrates. The conduction band offset at the $Al_2O_3$/GaN interface was calculated to be 2.13 eV, in agreement with theoretical predications. A non-zero field of 0.93 MV/cm in the oxide under flat-band conditions in the GaN was inferred, which we attribute to a fixed net positive charge density of magnitude $4.60 \times 10^{12}$ cm$^{-2}$ at the $Al_2O_3$/GaN interface. We provide hypotheses to explain the origin of this charge by analyzing the energy band line-up.

[a)] Author to whom correspondence should be addressed. Electronic mail: espostom@ece.osu.edu Tel: +1-614-688-8458

Insulators such as $SiO_2$, $SiN_x$, $Al_2O_3$, and $HfO_2$ have been proposed as a mean for gate current suppression and surface passivation of GaN metal insulator semiconductor high electron mobility transistors (MISHEMTs) [1-11]. Of these, $Al_2O_3$ has several desirable properties, including a large band gap, breakdown electric field, and ease of deposition. The performance of GaN-based MISHEMTs for mm-wave as well as power switching applications depends critically on the dielectric/GaN interface. This letter reports a quantitative analysis of the interface barrier of $Ni/Al_2O_3/GaN$ capacitors in terms of conduction band discontinuity, interface fixed charge and Fermi-level pinning effect.

The samples used in this experiment, as shown in Fig. 1(a), were grown using a Veeco rf-plasma molecular beam epitaxy system on semi-insulating GaN templates on sapphire from Lumilog, with threading dislocation density of $\sim 5 \times 10^8$ $cm^{-2}$. The epilayer consisted of 200 nm unintentionally doped GaN on substrate followed by 100 nm Silicon-doped GaN. The nominal doping density was $1 \times 10^{18}$ $cm^{-3}$. Figure 1(b) shows a scan of the as-grown surface taken using atomic force microscope in tapping mode. The oxide layers were deposited in a Picosun atomic layer deposition (ALD) system, using trimethylaluminum (TMA) and $H_2O$ as precursors. After a HCl-based removal of the excess gallium droplets on the surface, three different thicknesses – i.e. nominal 6 nm, 12 nm, and 18 nm – of oxide were deposited at 300°C on three different pieces cleaved out of the same sample. The pre-deposition treatment of the surface consisted in a 10:1 HF-dip for 15s. All three samples were then annealed at 600°C in forming gas for 1 min. The gate pads were defined by optical contact lithography. Features of large contacts were also defined in the photo resist. Buffered Oxide Etch (BOE) 10:1 was used to remove locally the oxide layer in these large features in order to get ohmic contacts. A Ni/Au/Ni stack was e-beam evaporated and a post metallization annealing was finally performed on all three samples at 400 °C in forming gas for 5 min.

C-V measurements were performed using an Agilent B1500 semiconductor device analyzer equipped with medium power source/monitor units (MPSMUs) and multi frequency capacitance measurement unit (MFCMU). At equilibrium, the GaN was found to be depleted at the surface for 6 nm and 12 nm ALD samples, and in accumulation for the 18nm thick ALD sample. Very low current densities below 10 nA/cm$^2$ over a wide voltage range, were measured for all three samples, indicating that the Al$_2$O$_3$ layers have excellent insulation properties.

The flat band voltage in the GaN for each of the structures was derived from the capacitance voltage profiles (Fig. 2), and were found to be 0.81 V, 0.21 V and -0.31 V for the 6 nm, 12 nm and 18 nm thick oxides respectively (inset to Fig. 2). The shift of the flat-band voltage as a function of oxide thickness is a clear indication of charges either at the interface or in the oxide. The experimental V$_{FB}$ vs. t$_{ox}$ data points plotted in the inset of Fig. 3, shows a linear relationship with a high degree of correlation.

We use energy band diagram analysis to understand the physical properties of the interface. Figure 3 shows a qualitative conduction band diagram of the MIS capacitors with different oxide thicknesses at flat-band. V$_{gi}$ is the flat-band gate bias for each thickness, t$_{oxi}$, φ$_b$ the barrier height at the Ni/Al$_2$O$_3$ interface, F$_{ox}$ the electric field in the oxide layer, ΔE$_c$ the conduction band discontinuity between Al$_2$O$_3$ and GaN, and φ$_s$ the energy separation of the conduction band from the Fermi level in the n$^+$ GaN layer. Assuming zero interfacial charge in the oxide, a simple analytical expression relating the applied flat band voltage to the interfacial parameters can be derived from Fig. 3 as

$$qV_{gi} = -qF_{ox}t_{oxi} + (\varphi_b - \Delta E_c - \varphi_s).$$

Here, $V_{gi}$ is the flat-band voltage for the given oxide thickness $t_{oxi}$. According to the analytical expression of the linear fit reported in the inset of Fig. 3, the electric field dropping across the oxide at flat-band is 0.93 MV/cm and the ($\varphi_b - \Delta E_c - \varphi_s$) band offset is 1.357 eV. Based on the doping density, the conduction band distance from the Fermi level ($\varphi_s$) is estimated to be 18 meV. Assuming a barrier height of 3.5 eV at the Ni/Al$_2$O$_3$ interface[12], the conduction band offset between Al$_2$O$_3$ and GaN is found to be 2.13 eV. This experimental value is in agreement with the theoretical prediction of 2.1 eV reported in Ref. [13]. The fixed Al$_2$O$_3$/GaN interface charge can be also estimated from the electric field in the oxide under semiconductor flat-band conditions. The non-zero field dropping across the oxide can be attributed to a net positive charge at the Al$_2$O$_3$/GaN interface, of approximately $4.60 \times 10^{12}$ cm$^{-2}$.

We note that since the Fermi level is nearly at the GaN conduction band edge, the positive states causing this cannot be attributed to GaN mid-gap donor states which would have to be neutral under these conditions. The presence of net positive interface charge can be explained by two possible scenarios. The first explanation [Fig. 4 (a)] is based on the presence of interfacial *fixed* charge, which we could attribute to energy states between the conduction band minima of Al$_2$O$_3$ and GaN. If we assume this, using the spontaneous GaN polarization charge ($\sigma_{sp\_GaN}$) of $1.81 \times 10^{13}$ cm$^{-2}$, we calculate the fixed interface charge ($\sigma_{oxide}$) density to be $2.27 \times 10^{13}$ cm$^{-2}$. Under both positive and negative bias on the MIS structure, the Fermi level cannot modulate these states, and they will therefore behave like "fixed" charges. We note that they can, however, be modulated by optical excitation, and UV investigation of MIS structures should provide very useful information on these interface charges.

An alternative explanation [Fig. 4 (b)] is based on the presence of Ga-O or Ga-Al bonds at the Al$_2$O$_3$/GaN interface, causing the inversion of the polarity of the surface. Such inversion of the polarity would result in the inversion of the polarization charge at the interface from negative to positive. Similar experimental observation is reported in Ref. [14], where an $AlO_x$ transition layer has been proven to be an effective mean for GaN polarity inversion. In such a case, we could explain the electrostatics of our system by assuming that a net positive spontaneous polarization charge of magnitude $4.60 \times 10^{12}$ terminates the surface. Further experiments are required to understand the origin of the electric field in the oxide.

In conclusion, a quantitative analysis of the $Al_2O_3$/GaN interface barrier is presented. Ni/$Al_2O_3$/GaN MIS capacitors were fabricated on plasma assisted molecular beam epitaxy grown GaN using ALD for the dielectric deposition. Very low current densities were observed for all three samples, proving the good insulating properties of the $Al_2O_3$ layers. A linear relationship between flat-band voltage and oxide thickness has been experimentally observed pointing out the absence of any Fermi-level pinning at the $Al_2O_3$/GaN interface, and the presence of interfacial charges. Assuming a Ni/$Al_2O_3$ barrier height of 3.5 eV, the conduction band offset between $Al_2O_3$ and GaN has been found to be 2.13 eV. This value well matches with the predicted value[13]. A net fixed interface $Al_2O_3$/GaN positive interface charge of $4.60 \times 10^{12}$ cm$^{-2}$ was shown to exist, and we provide two hypotheses to explain the origin of these charges.

**ACKNOWLEDGMENT**

The authors gratefully acknowledge the support of ONR-MURI DEFINE (Dr. Daniel S. Green) and the valuable help of Anisha Ramesh (OSU).

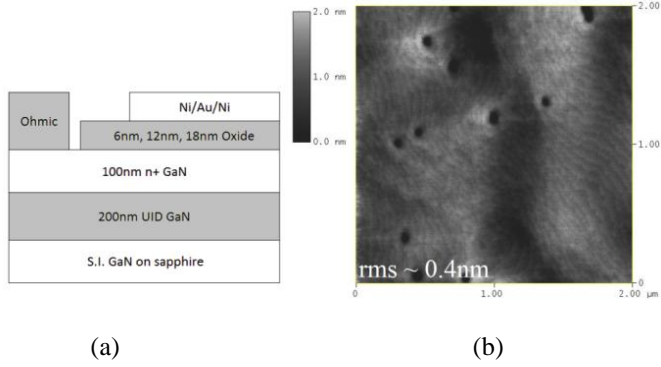

(a)                  (b)

Fig. 1 – (a) Schematic diagram of the MIS capacitors structure and (b) AFM image of the as grown surface.

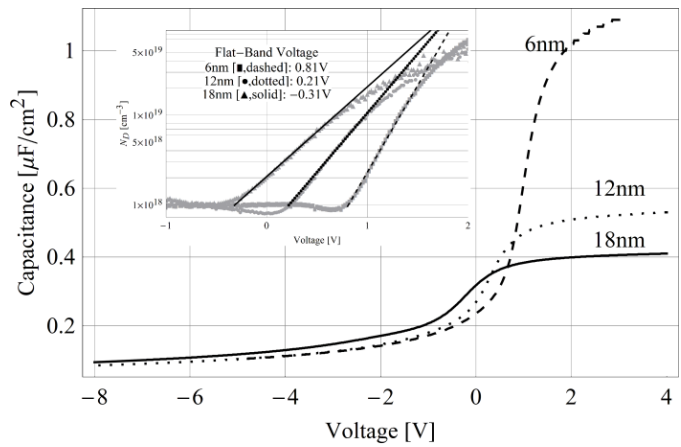

Fig. 2 – C-V characteristics of the MIS structures and (inset) extracted apparent charge profile as function of the applied bias. The exponential fitting of the accumulation regime and the flat-band voltages are also reported.

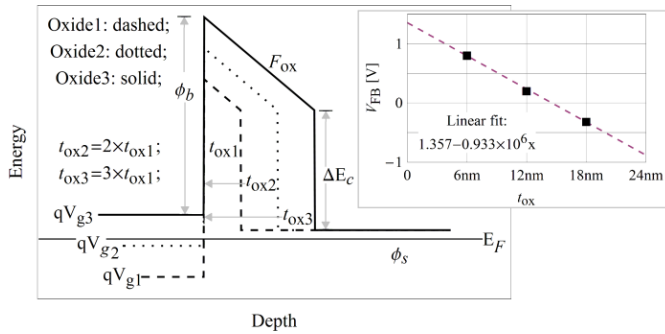

Fig. 3 – Schematic conduction band diagram of the MIS capacitors with different oxide thickness at flat-band and (inset) extraction of the Ni/$Al_2O_3$/GaN conduction band alignment and the oxide electric field $F_{ox}$ from the $V_{FB}$ vs. $t_{ox}$ plot..

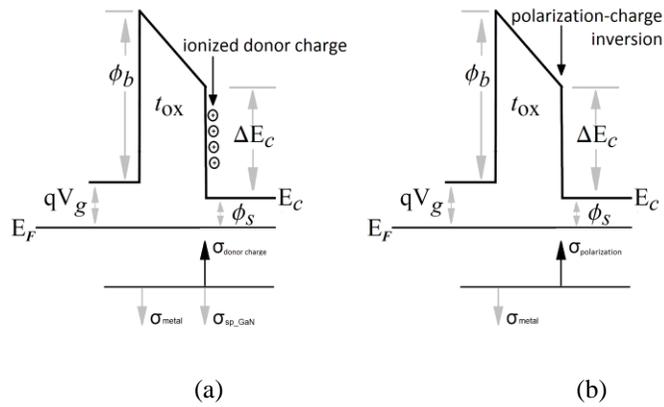

Fig. 4 – Qualitative energy band diagrams and charge distribution, assuming either (a) ionized donor charge or (b) polarization-charge inversion at the $Al_2O_3$/GaN interface.

…..